\documentclass[12pt]{article}
\usepackage{fancyhdr}

\usepackage{mathrsfs}
\usepackage[T1]{fontenc}
\usepackage{mathpazo}
\usepackage{setspace}
\usepackage{amsfonts}
\usepackage{amssymb}
\usepackage{amsmath}
\usepackage{epsfig}
\usepackage{latexsym}
\usepackage{color}
\usepackage{graphicx}
\usepackage{nicefrac}
\usepackage{mathtools}
\usepackage[latin1]{inputenc}
\usepackage{slashed}
\usepackage{multirow}
\usepackage{comment}
\usepackage{soul}
\usepackage{hyperref}
\usepackage[nosort]{cite}
\usepackage{datetime}
\usepackage{braket}
\newtheorem{theorem}{Theorem}

\usepackage[titletoc]{appendix}
\def\hybrid{\topmargin -20pt    \oddsidemargin 0pt
	\headheight 0pt \headsep 0pt
	\textwidth 6.25in       % A4 paper
	\textheight 9.25in       % A4 paper
	\marginparwidth .875in
	\parskip 5pt plus 1pt   \jot = 1.5ex}

\hybrid
\begin{document}
	\numberwithin{equation}{section}
	\begin{titlepage}
		\begin{center}
			\vskip .5 in

			{\large\bf Coherent-state path integrals in the continuum via geometric de-quantization}
			
			\vskip 0.35in
			
			{\bf P. Lykourgias}, {\bf I. Lyris}, {\bf A. I. Karanikas}
			\vskip 0.1in

			{\em{}Department of Nuclear and Particle Physics,\\ Faculty of Physics, National and Kapodistrian University of Athens,\\15784 Athens, Greece
			}

			\vskip 0.1in
			
			\vskip 0.1in

			{\footnotesize \texttt panoslyk@phys.uoa.gr, giannislyris@phys.uoa.gr, akaran@phys.uoa.gr}
			
			\vskip .43in
		\end{center}

		\centerline{\bf Abstract}
		\noindent
    
We present a new method for the consistent construction of time-continuous coherent-state path integrals using the theory of half-form quantization. Through the inversion of the quantization procedure we construct a de-quantization map taking first order operators to their corresponding path integrals. We generalize our results using functional techniques, allowing for the consistent path integral study of more general operators, including higher orders and interactions.

		\vskip .4in
		\noindent
	\end{titlepage}
	\vfill
	\eject
	\tableofcontents

	\def\baselinestretch{1.2}
	\baselineskip 20 pt
\section{Introduction} 
\interfootnotelinepenalty=10000

Since its birth, the most common way to describe quantum mechanics has been in terms of operators. This is in contrast to classical physics where all physical quantities are represented by functions. It is an undeniable fact, though, that calculations involving operators are in general much more complicated than those involving classical functions. This is because, when one deals with operators, ordering is very important; a product of operators, in general, acquires a different value upon changing the order its terms appear on it. This of course does not happen when one deals with a product of functions, since they always commute -or anticommute if they are Grassmann- with each other. These issues have been apparent since the earliest years of quantum mechanics, making the idea of some classical description of the theory very appealing. 

This was successfully addressed by Feynman via the path integral formalism\cite{feynman1, feynman2}, which became a milestone in modern physics. This formalism indeed manages to describe quantum systems in terms of classical functions. Feynman's work nowadays is a very important mathematical tool in many branches of physics, besides quantum mechanics, such as quantum field theory, statistical mechanics, quantum information theory and even polymer physics \cite{kleinert}. 

But the description of a quantum system in terms of classical objects comes with a price: an infinite number of integrations has to be performed, in principle. Indeed, in this formalism, usually the first object that arises is the discrete version of the path integral -which involves infinite integrations- and from this, its continuum version is obtained. Now, in the continuum there are some really useful identities which often allow for calculations to be performed analytically, thus bypassing the cumbersome -or in many cases impossible- task of calculating the infinitely many integrals pertaining to the discrete level. Thus, in order for the path integral to be practically useful, it is vital to define its continuum version consistently. As we shall see, this is not a trivial task\cite{problems1}-\cite{kara2}.

In this work, we are going to establish a prescription one should follow in order to rigorously define the path integral in the coherent state basis\cite{coherentpath2}-\cite{coherentpath6}. Coherent states are a very important class of states, with huge physical interest and they can also be employed to expand the idea of the path integral to a complex phase space, which in turn paves the road for a plethora of new possible applications. Unfortunately, when the standard route to define the path integral via these states is followed, inconsistencies have been reported, meaning that somewhere during the procedure there must be a pitfall. This paper aims to find a possible origin of these inconsistencies, to bypass this pitfall by proposing an inconsistency-free procedure and to give the limitations under which it can be followed.

The paper is organized as follows:

In Section 2 we briefly demonstrate the standard construction of the path integral, we discuss the aforementioned discrepancies and we briefly comment on a previous attempt to bypass them.

In Section 3 we analyze the method of geometric quantization, which we shall employ in order to establish our prescription. 

In Section 4 we construct a mapping from quantum operators to the classical functions which are appropriate for use in the continuum version of the path integral, based on the rigorous formalism that geometric quantization provides. We do that both for bosonic and spin systems.

In Section 5 we enlarge the set of operators on which the proposed mapping can act, encompassing specific higher order operators and interaction terms. 

The main text is accompanied by two Appendices, where we discuss mathematical details pertaining to the previous sections.

\section{Path integrals and coherent states}

First, we shall briefly outline the most common procedure that leads to the well-known path integral expressions\cite{kleinert}. We set $\hbar=1$ for simplicity and we consider one-dimesional, time-independent models. 

The transition amplitude from some state $\left| x_{a}\right\rangle$ at time $t_{a}$ to some state $\left| x_{b}\right\rangle $ at time $t_{b}$ is
\begin{equation} 
G=\left\langle x_b\right| \hat{U}(t_b, t_a)\left| x_a\right\rangle, \quad t_{a} <t_{b},  
\end{equation} 
where $\hat{U}(t_{b},t_{a})$ is the time evolution operator from time $t_{a}$ to time $t_{b}$. Then, the composition property of this operator is exploited in order to write it as a product of $N+1$ such operators, where each one of them expresses the evolution during a time interval of duration $\epsilon=\frac{t_{b}-t_{a}}{N+1}$. This is the so called slicing procedure, in the sense that time is sliced into smaller intervals. Afterwards -if the Hamiltonian is written in its usual form, i.e. as the sum of the kinetic energy and the potential- the identity written in terms of position eigenstates
\begin{equation}\label{2.3}
\hat{I}=\int_{-\infty}^{\infty}dx_j \left| x_j\right\rangle \left\langle x_j\right| ,\quad j=1, ...,N, 
\end{equation}
is inserted between all the slices, where $j$ labels the inserted identities. From this, after some algebra, including the insertion of another set of identities (one for each slice) written in terms of momentum eigenstates this time, the discrete version of the path integral is recovered
\begin{equation} 
G=\prod_{j=1}^{N}\int_{-\infty}^{\infty}dx_j\prod_{j=1}^{N+1}\int_{-\infty}^{\infty}\frac{dp_{j}}{2\pi}e^{i\sum_{j=1}^{N+1}(p_j(x_j-x_{j-1})-\epsilon H(p_j, x_j))},
\end{equation}
where $x$, $p$ denote position and momentum variables, respectively. If the quantum Hamiltonian of the system is written in terms of position and momentum operators and the potential depends solely on position, the mathematical procedure leading to the above relation reveals the classical function $H(p_j, x_j)$: it is just the quantum Hamiltonian with the operators $\hat{x}$, $\hat{p}$ replaced by classical functions $x$, $p$. 

Then, by taking the limit $\epsilon \rightarrow 0$, the continuum version of the path integral is acquired:
\begin{equation}
G=\int_{x(t_a)=x_a}^{x(t_b)=x_b}\mathcal{D}x\int \frac{\mathcal{D}p}{2\pi }e^{i\int_{t_a}^{t_b}dt(p(t)\dot{x}(t)-H(p(t), x(t)))}.
\end{equation}
Here, the symbol $\mathcal{D}a$ denotes functional integration with respect to the variable $a$ \cite{kleinert}. 

From this, one could also find a path integral expression for the partition function of a system, namely 
\begin{equation}
Z=\text{tr}\left[ e^{-i\int_{t_{a}}^{t_{b}}dt\hat{H}}\right] =\int_{pbc}\mathcal{D}x\int\frac{\mathcal{D}p}{2\pi}e^{i\int_{t_a}^{t_b}dt(p(t)\dot{x}(t)-H(p(t), x(t)))},
\end{equation}
where $pbc$ denotes periodic boundary conditions.   	

To check the validity of the path integral formalism, as an example, one could calculate the partition function of the harmonic oscillator (all units are dropped for simplicity)
\begin{equation}
\hat{H}=\frac{\hat{p}^{2}+\hat{x}^{2}}{2},\label{harmoscqu}
\end{equation}
which is an archetypal system in both classical and quantum physics. This quantum Hamiltonian indicates that the classical Hamiltonian symbol to insert in the action is
\begin{equation}
H=\frac{p^{2}+x^{2}}{2},
\end{equation}
leading to 
\begin{equation}
Z=\int_{pbc}\mathcal{D}x\int\frac{\mathcal{D}p}{2\pi}e^{i\int_{t_a}^{t_b}dt(p\dot{x}-\frac{p^{2}+x^{2}}{2})}. \label{partfunccorrect}
\end{equation}
The partition function can, then, be found via its path integral expression both from its discrete -albeit the calculation is rather tedious- and from its continuum version. At the same time, it can easily be obtained from its trace definition. Indeed, the discrete version of the path integral, the continuum one and the trace definition, all lead to the same result, which is the correct value 
\begin{equation}
Z=\sum_{n=0}^{\infty}e^{-iTE_{n}}=\sum_{n=0}^{\infty}e^{-iT(n+\frac{1}{2})}=\frac{1}{2isin\frac{T}{2}}, \quad T= t_b-t_a.
\end{equation}  

Now, it is not necessary for a quantum Hamiltonian to be expressed in terms of position and momentum operators specifically, while the partition function does not depend on the specific way a Hamiltonian is expressed, as can be seen by its trace definition. Thus, the same must be true for its path integral definition, as well. Trying to check this, one could write the quantum Hamiltonian (\ref{harmoscqu}) in terms of creation and annihilation operators. Then Eq. (\ref{harmoscqu}) becomes
\begin{equation}
\hat{H}=\hat{a}^{\dagger}\hat{a}+\frac{1}{2}, \label{harmosca}
\end{equation} 
where $\hat{a}^{\dagger}$, $\hat{a}$ are the creation and the annihilation operator, respectively. Now, upon trying to repeat the procedure that leads to the path integral, it becomes clear that another set of identities needs to be inserted between the slices. Indeed, the position and the momentum eigenstates are no longer very useful, since the Hamiltonian is not expressed in terms of position and momentum operators. Fortunately, there is a very convenient set of states which can be employed, namely the coherent states, 
\begin{equation}
\left| z\right\rangle=e^{-|z|^{2}/2}e^{z\hat{a}^{\dagger}}\left| 0\right\rangle =e^{-|z|^{2}/2}\sum_{n=0}^{\infty}\frac{z^{n}}{\sqrt{n!}}\left| n\right\rangle, \label{cohstate}
\end{equation} 
which are eigenstates of the annihilation operator with eigenvalue $z$. Note that they are not orthogonal to each other and that they can resolve the identity:
\begin{equation}
\hat{I}=\int_{\mathbb{C}}\frac{dzd\bar{z}}{2\pi i}\left| z\right\rangle \left\langle z\right|, \label{cohid}
\end{equation}
i.e. they span an overcomplete basis.

After this, the whole process can proceed the usual way, leading to the discrete version of the path integral for the transition amplitude from some coherent state $\left| z_{a}\right\rangle $ to another $\left| z_{b}\right\rangle $
\begin{multline}
G'=\left\langle z_{b}\right| e^{-iT \hat{H}}\left| z_{a}\right\rangle=\prod_{j=1}^{N}\Big(\int_{\mathbb{C}}\frac{d\bar{z}_{j}dz_{j}}{2\pi i}\Big)e^{-\sum_{j=1}^{N+1}(\frac{1}{2}[\bar{z}_{j}(z_{j}-z_{j-1})-(\bar{z}_{j}-\bar{z}_{j-1})z_{j-1}]+i\epsilon H_{j})},
\end{multline}
where $\frac{\left\langle z_{j}\right| \hat{H}\left| z_{j-1}\right\rangle}{\left\langle z_{j}| z_{j-1}\right\rangle}=H_{j}$. 
Note that now the expression is written in terms of complex variables. Once again, the limit $\epsilon \rightarrow 0$ gives the continuum version of the transition amplitude
\begin{equation}
G'=\int_{z(0)=z_{a}}^{\bar{z}(T)=\bar{z}_{b}}\mathcal{D}^{2}ze^{\frac{\bar{z}_{b}z(T)+\bar{z}(0)z_a-|z_a|^{2}-|z_{b}|^{2}}{2}}e^{i\int_{0}^{T}dt[i\frac{\bar{z}\dot{z}-\dot{\bar{z}}z}{2}-H(z, \bar{z})]}
\end{equation}
where the irrelevant factors $2\pi i$ have been absorbed into the integration measure. Then, after enforcing periodic boundary conditions $\left| z_{a}\right\rangle =\left| z_{b}\right\rangle =\left| z'\right\rangle$  and integrating over the boundary variables $z'$, $\bar{z}'$ the partition function written in terms of complex variables is recovered:
\begin{equation}
Z'=\int_{pbc}\mathcal{D}^{2}ze^{i\int_{0}^{T}dt[i\frac{\bar{z}\dot{z}-\dot{\bar{z}}z}{2}-H(z, \bar{z})]}. \label{partfuncz0}
\end{equation}
Now, for the harmonic oscillator, the classical function $H(z, \bar{z})$ can be read from the expression for $H_{j}$ after taking the limit $\epsilon \rightarrow 0$, which gives
\begin{equation}
H=|z|^{2}+\frac{1}{2},
\end{equation}
meaning
\begin{equation}
Z'=e^{-\frac{iT}{2}}\int_{pbc}\mathcal{D}^{2}ze^{i\int_{0}^{T}dt[i\frac{\bar{z}\dot{z}-\dot{\bar{z}}z}{2}-|z|^{2}]}. \label{partfuncz}
\end{equation}
This result, though, is not the correct one. Indeed, it has already been established that Eq. (\ref{partfunccorrect}) gives the correct result. Then, by performing the canonical transformation $x=\frac{z+\bar{z}}{\sqrt{2}}$, $p=i\frac{\bar{z}-z}{\sqrt{2}}$, Eq. (\ref{partfunccorrect}) gives
\begin{equation}
Z=\int_{pbc}\mathcal{D}^{2}ze^{i\int_{0}^{T}dt[i\frac{\bar{z}\dot{z}-\dot{\bar{z}}z}{2}-|z|^{2}]}.
\end{equation}
This means that the calculation of the quantity in Eq. (\ref{partfuncz}), i.e. the calculation via the path integral in the coherent states basis, leads to the incorrect result $Z'=\sum_{n=0}^{\infty}e^{-iT(n+1)}=e^{-\frac{iT}{2}}Z$. 

In fact, such incosistencies do not pertain solely to bosonic systems. Considering spin $s$ systems and repeating the usual procedure, the most natural resolution of the identity would be in terms of spin coherent states\cite{coherentpath9+identitycoherentidentity3+spincalc+continuumsolari1, spindef, coherentidentity4+spincalc+extraphase}
\begin{equation}
\hat{I}=\frac{2s+1}{2\pi i}\int_{\mathbb{C}}\frac{d\bar{z}dz}{(1+|z|^{2})^{2}}\ket{z}_{s}\prescript{}{s}{\bra{z}}, \label{spinid}
\end{equation}
where 
\begin{multline}
\left| z\right\rangle_{s} =\frac{1}{(1+|z|^{2})^{s}}e^{z\hat{S}_{-}}\left| s,s\right\rangle=\frac{1}{(1+|z|^{2})^{s}}\sum_{j=-s}^{s}\Big[\frac{(2s)!}{(s-j)!(s+j)!}\Big]^{1/2}z^{s-j}\left| s,j\right\rangle \label{spincohstate}
\end{multline}
are the spin coherent states and $\left| s,j\right\rangle $ denotes the eigenstate of $\hat{{S}}_{z}$ with eigenvalue $j$. Then, considering even the simplest possible spin Hamiltonian $\hat{H}=\omega\hat{S}_{z}$ and following the standard procedure, we come to the wrong conclusion that for this Hamiltonian\cite{kara2}
\begin{equation}
Z^{(s)'}=e^{\frac{i\omega T}{2}}\sum_{j=-s}^{s}e^{-i\omega Tj}=e^{\frac{i\omega T}{2}}Z^{(s)}.
\end{equation}
To make matters even worse, when calculations are performed in the coherent states basis, we encounter the unacceptable fact that the continuum result appears to be very sensitive to the discretization procedure. To see this, consider again the Hamiltonian in Eq. (\ref{harmosca}). After slicing the time evolution, inserting identities between the slices is not the only way to proceed. Alternatively, the slices can be manipulated in a symmetric way, namely 

\begin{equation}
e^{-i\epsilon\hat{H}}=\frac{1}{2\pi{i}}\int\left|z_j\right\rangle e^{-i\epsilon H(z_{j}, \bar{z}_{j})}d\bar{z}_j dz_j\left\langle z_j\right|, \label{alterslice} 
\end{equation}
where $H(z_{j}, \bar{z}_{j})$ needs to be found. This leads to

\begin{equation}
\left\langle z_{k}\right| e^{-i\epsilon\hat{H}}\left| z_{k}\right\rangle =\frac{1}{2\pi{i}}\int\left\langle z_{k}| z_{j}\right\rangle e^{-i\epsilon H(z_{j},\bar{z}_{j})}d\bar{z}_j dz_j\left\langle  z_{j}|z_{k}\right\rangle. 
\end{equation} 

Exploiting expression (\ref{cohstate}), and using the ansatz $H(z_{j},\bar{z}_{j})=A|z_{j}|^{2}+Bz_j+C\bar{z}_j+D$, it is easy to find that $H(z_{j},\bar{z}_{j})=|z_{j}|^{2}-\frac{1}{2}$. Then, substituting Eq. (\ref{alterslice}) in the expression for the transition amplitude in terms of coherent states and taking the limit $\epsilon \rightarrow 0$, once again an expression for the transition amplitude in the continuum is obtained, which can be used to acquire the partition function. Indeed, Eq. (\ref{partfuncz0}) is recovered, only this time $H=|z|^{2}-\frac{1}{2}$, which, in the same fashion as above, would attribute to the partition function an extra factor $e^{\frac{iT}{2}}$. 

This shows that different slicing procedures lead to different results in the continuum. Specifically in the case we studied above, both of them are wrong. Note, though, that the kinetic term was the same in both procedures -and the correct one, as well. This is due to the fact that it stems from the inner product between neighboring states. Thus, the only problematic term was the Hamiltonian symbol. Note, also, that if the discrete version of the path integral is used for the calculation, the correct result is always reproduced, no matter how the slices were manipulated.

There have been some attempts to fix these issues\cite{kara1, kara2, coherentpath9+identitycoherentidentity3+spincalc+continuumsolari1, continuumsolari2+spincalc}. For example, in  \cite{kara1} the authors present a specific way to acquire the classical Hamiltonian in the continuum. The validity of the procedure was confirmed, but only in a few specific cases. Nevertheless, even if such a recipe could be proven to be valid in any instance, it does not really resolve the issues concerning the coherent state path integral; it bypasses them in a specific way. 

Such incosistencies have been reported only when overcomplete sets of states, such as the coherent states, are used during the slicing manipulation. On the other hand, if orthonormal states are used  during the construction, the continuum limit appears to be uniquely defined under the aforementioned assumptions regarding the form of the Hamiltonian \cite{kleinert}. This indicates that when one tries to define the continuum version of the path integral, they need to be very careful during the discretization procedure, since incosistencies might appear at that stage. The inconsistencies were apparent here, because the harmonic oscillator is a system that has been studied thoroughly, so the anticipated results are already known. Not all systems share such a property, though. It must, also, be noted that the results pertaining to the simple systems considered here are correct up to a phase, while for more complicated systems, if the usual procedure is used, an extra phase is not enough to fix the discrepancies.

As stated above, the continuum version of the path integral is very important in order for it to be a practical mathematical tool. At the same time, the above analysis indicates that the usual procedure, from the discrete version to the continuum, is not always trustworthy. This observation, calls for a recipe that could determine the classical Hamiltonian symbol in the action on the continuum without reference to the discrete. The establishment of such a recipe is the goal of the following Sections. Key role to this has the kinetic term, which stems from the inner product between states and is always well defined.

\section{Elements of geometric quantization}

To put the problem on a firm ground, keeping in mind the above analysis which indicates that we need to bypass the slicing procedure, we are looking for a mapping that takes a quantum Hamiltonian, i.e. an operator, and maps it to a classical Hamiltonian, i.e. a function. As explained in Section 2, naively considering as classical Hamiltonian the expectation value of the quantum one with respect to a set of coherent states does not amount to a consistent map. For example for the harmonic oscillator Hamiltonian $\hat{H}=\hat{a}^{\dagger}\hat{a}+\frac{1}{2}$, this gives $H=|z|^{2}+\frac{1}{2}$, which leads to incosistencies. 

Actually, the pursuit of a correspondence between classical functions and operators is not a new idea at all. Dirac was the first one that established the constraints such a mapping has to obey in order to be consistent, albeit he had in mind an inverse correspondence than the one we need here, i.e. a mapping from classical functions to quantum operators. Unfortunately, it was later proven that such a mapping cannot exist, but only for a limited set of functions\cite{obstruction, groenewolds2}. This was the verdict of the Groenewold-van Hove theorem, which we present along with its proof in Appendix A. 

Moreover, there is a deeper theoretical reason which indicates that such a task is not trivial at all: there are great conceptual and structural differences between functions and operators. In terms of geometry, functions lie on a phase space which turns out to be a symplectic manifold, while operators act on a Hilbert space, a structure defined over the aforementioned manifold. Geometric quantization is a procedure that provides such a correspondence, in the way Dirac intended, i.e. functions are mapped to operators. Below we give the basic ingredients of this method. The mathematical concepts it relies on are rather technical and profound, so, in order to understand their meaning, we shall try to give their significance in our context. 

The Hilbert space over curved symplectic manifolds is usually considered to be spanned by sections of a complex line bundle, where the connection is identified with the symplectic potential $A$. A complex line bundle is a structure which is constructed upon choosing a complex 1D vector space for every point over a topological manifold, in a smooth manner. Sections are locally identified as functions $\Phi$ on a coordinate chart, while connection is an object used to define parallel transport over a manifold in a consistent way. Unfortunately, in the context of geometric quantization, which remains the only consistent quantization method for curved manifolds, this formalism does not always suffice for the construction of real irreducible representations of quantum mechanics. 

The solution to this was given through the redefinition of the Hilbert space as the set spanned by tensor products of sections of the previous complex line bundle with sections of the half-form bundle, where the elements of the latter behave locally as square roots of differential forms. The symplectic potential was again chosen to be the connection of the new structure. This formalism turned out to be consistent and can explicitly construct irreducible representations over any classical, real or complex, manifold. Thus, in order for the representation of quantum mechanics to be irreducible for both real and complex manifolds, we consider the, so called, half -form quantization.

In our context, the manifolds we are interested in, are those defined by the phase space of the path integral. More specifically, we refer to complex manifolds, since these are the ones where incosistencies were encountered. In literature, the wave functions over such manifolds are usually defined as the set of holomorphically polarized complex line sections. This polarization is defined through the action of the anti-holomorphic covariant derivative
\begin{equation}
D_{\bar{z}}\Phi(z,\bar{z})=0\implies\left(\frac{\partial}{\partial\bar{z}}+iA_{\bar{z}}\right)\Phi(z,\bar{z})=0, \label{polar}
\end{equation}
where the connection $A$ is the symplectic potential on the underlying symplectic manifold $M$. This equation has a solution of the form $\Phi(z,\bar{z})=\phi(z)\text{exp}\left[-{Y(z,\bar{z})/2}\right]$, where $Y$ is a function for which $\text{Re}\left[Y(z,\bar{z})\right]\rightarrow+\infty$ sufficiently fast in the $|z|\rightarrow+\infty$ limit. Then, a canonical inner product can be defined \cite{qftmath} for the wave functions as
\begin{equation}
(\Phi_1,\Phi_2)=k\int_M\phi_1^*(\bar{z})\phi_2(z)e^{-Y(z,\bar{z})}\mu(z,\bar{z}),\label{resold}
\end{equation}
where $k$ is a normalization constant and $\mu(z,\bar{z})$ is the measure of integration defined canonically on the 2D manifold $M$. In this case, this is simply the symplectic 2-form\footnote{In the general case of 2n dimensions the canonical measure is given as
\begin{equation}
	\mu(x^i)=\frac{\omega^n}{n!}=\sqrt{\text{det}\omega}dx^1\wedge{d}x^2\wedge\cdots\wedge{d}x^{2n}\nonumber,
	\end{equation}
	where the $x^i$ 's with $i=1,2,\cdots,2n$ are the coordinates of a local chart.} $\omega=-dA$, i.e. $\mu(z,\bar{z})=\omega$. In principle, the symplectic 2-form is the fundamental quantity defining the symplectic structure, but in the context of geometric quantization the symplectic potential has a more prominent role. 

In order for the method of quantization to have the same structure over both real and complex symplectic manifolds, one, also, has to promote the set of holomorphically polarized complex-line sections to a set of holomorphically polarized half-form corrected ones \cite{geom1, woodhouse, qftmath, geom2}. In this formalism, the holomorphically polarized wave-functions are locally of the form
\begin{equation}
\Phi(z,\bar{z})=\phi(z)e^{-\frac{Y(z,\bar{z})}{2}}\sqrt{dz}, \label{halfform}
\end{equation} 
where the polarization of the function part remains the same as before and the condition of holomorphic polarization for the half-form part is also taken into account.
The canonical inner product is then defined as \cite{qftmath}\footnote{In what follows, the symbol $\frac{a}{b}$ with $a$ and $b$ being differential forms of the same degree, is the function for which $\left(\frac{a}{b}\right)\times{b}=a$}
\begin{equation}
(\Phi_1,\Phi_2)=k\int_M\left(\frac{(\Phi_1^*\otimes\Phi_1^*)\wedge(\Phi_2\otimes\Phi_2)}{\mu(z,\bar{z})}\right)^\frac{1}{2}\mu(z,\bar{z}), \label{innerproduct}
\end{equation}
where $k$ again is a normalization constant.

By substituting Eq. (\ref{halfform}) in (\ref{innerproduct}) this expression simplifies to
\begin{equation}
(\Phi_1,\Phi_2)=k\int_M\phi_1^*(\bar{z})\phi_2(z)e^{-Y(z,\bar{z})}\left(\frac{dz\wedge{d\bar{z}}}{\mu(z,\bar{z})}\right)^\frac{1}{2}\mu(z,\bar{z}). \label{innerkahler}
\end{equation}

The above reasoning may seem rather technical. Consider, though, its main points: we begin with a symplectic manifold. Then, the irreducible representations of wave functions are defined in a way that depends on the connection. Finally, for them we define a canonical inner product or, equivalently, a Hilbert space has been constructed. In other words, from a symplectic manifold -the phase space- emerges a Hilbert space -where wave functions reside. Now, it remains to find a correspondence between the objects relevant to each one of them. These are classical functions and operators, respectively, hence the relevance of this method to our scope. 

Regarding the observable, classical functions, geometric quantization proceeds by identifying, for each one of them, the differential (quantum) operator \cite{geom2}, the observable effect of which is represented by the classical function at hand. For a smooth function $f$, considered as a classical observable on a manifold, the operator constructed through the half-form extended geometric quantization procedure \cite{woodhouse, qftmath} is 
\begin{equation}
\hat{Q}(f)=-i\xi^{z}\partial_{z}-\frac{i}{2}\left(\partial_{z}+4iA_{z}\right)\xi^{z}+f, \label{operator}
\end{equation}
where $\xi^\mu=\omega^{\mu\nu}\partial_\nu{f}$ is the Hamiltonian vector field corresponding to the smooth function $f$. In Appendix B we present the basic steps needed to arrive at this result.
In this form, operator (\ref{operator}) acts only on the holomorphic function $\phi(z)$ of the half-form corrected polarized wave functions (\ref{halfform}).  The contribution of the half-form correction is included in the $\partial_{z}\xi^{z}$ term and is known as the metaplectic correction \cite{geom1, woodhouse, qftmath, geom2}. This term will prove to be crucial for the cure of the incosistnecies encountered above, as we shall see in the next Section.

A very strict constraint this construction must adhere to, refers to the Hamiltonian vector field $\xi$, the flow of which induces the action of the respective operator. This field must preserve the polarization \cite{geom1, woodhouse, qftmath, geom2}, i.e. its Lie Derivative acting on any polarized vector field $P=\phi(z)\frac{\partial}{\partial{z}}$ must lie in the space of polarized vector fields
\begin{equation}
\mathcal{L}_\xi\left(P\right)=[\xi,P]=\left(\xi^z\partial_z\phi-\phi\partial_z\xi^{z}\right)\partial_{{z}}\quad\forall{z}\in\mathbb{C}, \label{polarize}
\end{equation}
with the coefficient of the antiholomorphic direction $-\phi\partial_z\xi^{\bar{z}}$ being equal to zero. Thus, it suffices to have 
\begin{equation}
\partial_z\xi^{\bar{z}}=0.\label{better}
\end{equation}
If this condition is not met, the operator is not legitimate, since the preservation of the polarization is a strict condition for the validity of the whole procedure. It must, also, be noted that the procedure leading to Eq. (\ref{operator}) is expected to hold only for operators which contain derivatives up to the first order, since only these are related linearly to vector fields. In Section 5 we argue that, in the path integral context, the extension of this procedure to higher powers of such operators, can be performed with the use of functional techniques, without the introduction of extra mathematical structure.

The main idea in the current work, advocated in the next section, is to consider the quantum operator as known and use Eq. (\ref{operator}) as a first order differential equation with respect to the function $f$. In that case, we demand that the condition (\ref{better}) is fulfilled by the Hamiltonian vector field corresponding to the function $f$, regardless if the metaplectic correction is taken into account or not. This way we aim to understand how the action of that operator is represented as a classical observable on a given symplectic manifold and use this reasoning to identify the correct Hamiltonian symbol, weighing time-continuous path integrals.

\section{Path integral construction in the continuum}

In this Section we will showcase how half-form quantization can be used to identify the correct continuum limit for the Hamiltonian symbol appearing in path integrals. To connect geometric quantization with the usual Dirac bra/ket notation we express the set of coherent states $\ket{z}$ through their more abstract mathematical definition \cite{qftmath}. These are elements of the Hilbert space, the inner product of which with an arbitrary state $\ket{\phi}$ gives the value of the corresponding holomorphic wavefunction at a point $z$ 
\begin{equation}
\phi(z)=\braket{z|\phi}.
\end{equation}
This definition allows Eq. (\ref{innerkahler}) to be rewritten as
\begin{equation}
(\Phi_1,\Phi_2)=\bra{\phi_1}\left\{k\int_M\ket{z}\bra{z}e^{-Y(z,\bar{z})}\left(\frac{dz\wedge{d\bar{z}}}{\mu(z,\bar{z})}\right)^\frac{1}{2}\mu(z,\bar{z})\right\}\ket{\phi_2},
\end{equation}
through which the resolution of the identity can be formally defined 
\begin{equation}
\hat{I}=k\int_M\ket{z}\bra{z}e^{-Y(z,\bar{z})}\left(\frac{dz\wedge{d\bar{z}}}{\mu(z,\bar{z})}\right)^\frac{1}{2}\mu(z,\bar{z}). \label{identity}
\end{equation}
The existence of the resolution of the identity\footnote{In the case where the the half-form structure is not included, the same argument used on Eq. (\ref{resold}) gives again a resolution of the identity
	\begin{equation}
	\hat{I}=k\int_M\ket{z}\bra{z}e^{-Y(z,\bar{z})}\mu(z,\bar{z})\nonumber. 
	\end{equation}} 
allows for the use of coherent states during the discretization of -real or imaginary- time evolution, through which the path integral representation of the transition amplitude, the generating functional or the partition function is constructed \cite{kleinert}\cite{pathcoherent}-\cite{pathspincoherent2}.
It is easy to see that both Eqs. (\ref{cohid}) and (\ref{spinid}) reduce to the form of Eq. (\ref{identity}). To proceed, we define the holomorphic coherent states by dropping the normalization factors from the expressions in Eq. (\ref{cohstate}), (\ref{spincohstate}) which in turn are absorbed into the identities (\ref{cohid}), (\ref{spinid}). Thus, for a bosonic system we consider the holomorphic coherent states
\begin{equation}
\ket{z}_b^{(0)}=e^{z\hat{a}^\dagger}\ket{0}=\sum_{n=0}^{\infty}\frac{z^n}{\sqrt{n!}}\ket{n}, \label{coboson}
\end{equation}
with resolution of the identity \cite{coherentpath7+identity, coherentpath9+identitycoherentidentity3+spincalc+continuumsolari1, coherentpath8+identity}
\begin{equation}\label{identbos}
\hat{I}_{E_2}=\frac{1}{2\pi{i}}\int_\mathbb{C}\ket{z}_{b}^{(0)}\prescript{(0)}{b}{\bra{z}}e^{-|z|^2}dz\wedge{d\bar{z}},
\end{equation}
(where the upper index $(0)$ denotes the lack of normalization) and for a spin system in the $su(2)$ representation with highest weight $s$ we consider the holomorphic coherent states
\begin{equation}
\ket{z}_s^{(0)}=e^{z\hat{S}_-}\ket{s,s}=\sum_{j=-s}^{s}\left[\frac{(2s)!}{(s-j)!(s+j)!}\right]^\frac{1}{2}z^{s-j}\ket{s,j}, \label{cospin}
\end{equation}
with the resolution of the identity \cite{coherentpath9+identitycoherentidentity3+spincalc+continuumsolari1, coherentidentity4+spincalc+extraphase}
\begin{equation}
\hat{I}_{S_2}=\frac{2s+1}{2\pi{i}}\int_\mathbb{C}\ket{z}_{s}^{(0)}\prescript{(0)}{s}{\bra{z}}e^{-2s\text{ln}[1+|z|^2]}\frac{dz\wedge{d\bar{z}}}{(1+|z|^2)^2}.\quad \label{identspin}
\end{equation}
From the form of Eqs. (\ref{identbos}) and (\ref{identspin}) it is easy to see that the symplectic manifolds over which the corresponding quantum mechanics are defined are the 2D Euclidean plane for the bosonic case, and the 2-Sphere for the spin case. In the following, we will denote the measure as $d^2z=dz\wedge{}d\bar{z}$, in order to simplify the integral expressions.

The discrete construction of a path integral over the set of the aforementioned coherent states is well defined, as long as it is considered as a product of countably infinite terms. For a general slicing procedure, the discrete version of the path integral partition function is given by 
\begin{equation}
\text{tr}\left[\hat{T}e^{-i\int_{0}^{T}dt\hat{H}}\right]\label{discreet}=
\lim_{\epsilon\rightarrow0}\lim_{{N\rightarrow\infty}}\underset{\text{pbc}}{\int}\prod_{j=0}^{N}\mu({z}_j,\bar{z}_j)\prod_{j=0}^{N}\braket{{z}_{j+1}|{z}_j}e^{-i\epsilon H_j},
\end{equation} 
where $\ket{z_{N+1}}=\ket{z_0}$ is the periodicity condition and $H_j$ is the classical Hamiltonian at time $t=\epsilon{j}$. In our approach, we demand for the limit of the aforementioned countably infinite product to an uncountably infinite one to be well defined in the context of Hamiltonian mechanics, a demand which is vital for the existence of a consistent time-continuous path integral. Next, we quantify this demand by requiring the continuum version of the path integral to coincide with the mathematically well-defined phase space path integral originating from the theory of geometric quantization. The key ingredient in our proposal is the identification of the classical Hamiltonian weighing the path integral with a specific function, namely the one which, upon half-form quantization, is mapped to the quantum Hamiltonian the path integral pertains to. In the light of the arguments presented in Section 3, this is a natural demand, corroborated by the mathematically strict correspondence between functions and operators, presented there. The continuum version of the path integral for the partition function is then defined to be of the form \cite{woodhouse, local}

	\begin{equation}
	\text{tr}\left[\hat{T}e^{-i\int_{0}^{T}dt\hat{H}}\right]= \label{pathintegral} \underset{\text{Periodic}}{\int}\mathcal{N}\left[\prod_{t\in[0,T]}d^2z(t)\sqrt{\text{det}||\omega(z(t),\bar{z}(t))||}\right]e^{i\int_{0}^{T}dt\left\{{A}_\mu(z(t),\bar{z}(t))\dot{x}^\mu(t)-H(z(t),\bar{z}(t))\right\}},
	\end{equation}

where $A_\mu$ is the symplectic potential of the manifold, on which the classical action is defined on, $x^{\mu}=\{z,\bar{z}\}$ is the set of complex coordinates, and

\begin{equation}
\prod_{t\in[0,T]}d^2z(t)\sqrt{\text{det}||\omega(z(t),\bar{z}(t))||}=\underset{N\rightarrow\infty}{lim}\prod_{j=0}^{N}d^2z_j\sqrt{\text{det}\omega(z_j,\bar{z}_j)} \label{measure}
\end{equation} is the functional measure. The Hamiltonian function $H$ in this formula represents the classical observable that controls, through the flow of the corresponding vector field, the evolution of the wave-functions on the Hilbert space. At this point it is important to note that the discrete structure that supports the integral (\ref{pathintegral}) is symmetric by construction.

The ''normalization factor'' $\mathcal{N}$ stems from the fact that the measure of integration $\prod_{t\in[0,T]}\mu(z(t),\bar{z}(t))=\underset{N\rightarrow\infty}{lim}\prod_{j=0}^{N}\mu(z_j,\bar{z}_j)$, originating from the discrete version, may not match perfectly with the symplectic structure appearing in the action\footnote{In the sense that $\mu(z,\bar{z})\neq\omega(z,\bar{z})=-dA$.} and can be modified as
\begin{align*}
\mu(z(t),\bar{z}(t))&=\frac{\mu(z(t),\bar{z}(t))}{\sqrt{\text{det}||\omega(z(t),\bar{z}(t))||}dz(t)\wedge{}d\bar{z}(t)}\sqrt{\text{det}||\omega(z(t),\bar{z}(t))||}dz(t)\wedge{}d\bar{z}(t)\equiv\\
&\equiv\mathcal{N}\sqrt{\text{det}||\omega(z(t),\bar{z}(t))||}dz(t)\wedge{}d\bar{z}(t)
\end{align*} 
in order to bring forth the canonical measure of the symplectic manifold. This matching becomes necessary when one wants to perform semi-classical calculations at the continuum limit.
If this factor is constant, it can just be factored out, as in the case of bosonic and spin systems \cite{local} where 
\begin{equation}
\mathcal{N}_b=\underset{N\rightarrow\infty}{lim}\prod_{j=0}^{N}\frac{-1}{2\pi},\quad\mathcal{N}_S=\underset{N\rightarrow\infty}{lim}\prod_{j=0}^{N}\frac{s+1/2}{s}\frac{-1}{2\pi}.
\end{equation}

Otherwise, it must be implemented to the action as a functional determinant similarly to the Faddeev-Poppov procedure of non-Abelian Gauge theories \cite{fadeev}.  Formula (\ref{pathintegral}) will be our definition of a time-continuous path integral in the coherent-state basis. 

The first step in our construction, before changing our point of view to geometric quantization, is the identification of the manifold over which path integration takes place. The symplectic potential ${A}$, defining the underlying symplectic structure in the continuum limit (\ref{pathintegral}), can be easily found from the limit of the countably infinite product of inner product terms in the discrete version of the path integral as
\begin{equation}
\lim_{\epsilon\rightarrow0}\lim_{N\rightarrow\infty}\prod_{j=0}^{N}\braket{{z}_{j+1}|{z}_j}\equiv{}e^{i\int_{0}^{T}dt{A}_\mu(z(t),\bar{z}(t))\dot{x}^\mu(t)}. \label{limit0}
\end{equation}
This limit can be easily seen to be well defined, since it is independent of the slicing manipulation. At the same time, the symplectic potential ${A}$ can be interpreted as an emergent connection on the curved classical manifold, since it dictates how neighboring time slices are sewed together at the continuum limit.

Due to the interpretation of the continuum version of the path integral as Eq. (\ref{pathintegral}), the limit of the Hamiltonian term is now expected to represent the classical observable on the manifold with symplectic potential $A$. Formally, for operators represented as first order differential ones and which correspond to polarization preserving Hamiltonian vector fields, this limit should be the smooth function $f$ satisfying Eq. (\ref{operator}). 

To proceed, we define the classical function appearing in Eq. (\ref{operator}) as the inverse of $\hat{Q}$ acting on the operator $\hat{H}$, while the operators for which this procedure is valid are defined as $\hat{Q}^{-1}$-de-quantizable. We can then explicitly define the continuum limit of the Hamiltonian symbol for the set of $\hat{Q}^{-1}$-de-quantizable operators (labelled by the index $d.q.$) as ${\hat{Q}}^{-1}(\hat{H}_{d.q.})(z(t),\bar{z}(t))$
{\begin{equation}
	\lim_{\epsilon\rightarrow0}\lim_{N\rightarrow\infty}H_{j,d.q.}\equiv{\hat{Q}}^{-1}(\hat{H}_{d.q.})(z(t),\bar{z}(t)). \label{limit}
	\end{equation}
	It must be noted that if Eq. (\ref{operator}) is to be used, the representation of an operator as a differential one emerges from its action on the purely holomorphic form of the coherent states \ref{coboson}, \ref{cospin}, i.e. when these are not normalized. We stress the fact that this operator representation is not expected to be a representation in the Lie-algebra sense, since the coordinate induced forms these operators take are just the result of their specific action on the corresponding set of coherent states. It is also easy to see that the Hamiltonian symbol corresponding to the identity operator is trivially $\hat{Q}^{-1}(\hat{I})=1$ over all symplectic manifolds.
	
	This procedure is not expected to be valid if the Hamiltonian vector field corresponding to an operator does not preserve the polarization, i.e. for operators that do not belong in the set of $\hat{Q}^{-1}$-de-quantizable ones. 
	
	We now proceed with the study of the bosonic and the spin case.
	
	\subsection{Bosonic coherent states}
	
	The identification (\ref{limit0}) in the basis of coherent states (\ref{coboson}) yields the following result
	\begin{equation}
	A_\mu\dot{x}^\mu=\frac{i}{2}{(\bar{z}\dot{z}-z\dot{\bar{z}})}.
	\end{equation}
	From this, the symplectic potential of the induced classical mechanics, corresponding to the 2D Euclidean plane, is found to be ${A}=\frac{i}{2}(\bar{z}dz-zd\bar{z})$.
	The coordinate induced forms for the annihilation and creation operators on the flat manifold are 
	\begin{equation}
	\hat{a}\ket{z}_b^{(0)}{=}z\ket{z}_b^{(0)}
	\quad\text{and}\quad
	\hat{a}^\dagger\ket{z}_b^{(0)}{=}\frac{\partial}{\partial{z}}\ket{z}_b^{(0)}
	\end{equation} respectively, while for the number operator we find\footnote{These forms can be found explicitly by acting with the afforementioned operators on the set of coherent states, and as explained previously do not constitute a representation of the algebra.}
	\begin{equation}
	\hat{a}^\dagger\hat{a}\ket{z}_b^{(0)}{=}z\frac{\partial}{\partial{z}}\ket{z}_b^{(0)}.
	\end{equation}
	Plugging these into Eq. (\ref{operator}) we find the respective smooth functions
	\begin{equation}
	\hat{Q}^{-1}(z)=z,\quad{\hat{Q}^{-1}\left(\frac{\partial}{\partial{z}}\right)}=\bar{z} \label{bosonz}
	\end{equation}
	and
	\begin{equation}
	\hat{Q}^{-1}\left(z\frac{\partial}{\partial{z}}\right)=|z|^2-\frac{1}{2}, \label{bosonzz}
	\end{equation}
	which indeed are known to provide correct results \cite{kleinert, kara1} when used as classical Hamiltonians in the respective time-continuous coherent state path integrals. In all considerations a symmetric underlying discrete structure is implied\footnote{Quantum field theory methods based on the computation of a functional determinant necessarily imply a symmetric underlying discrete structure.}, that is $|z|^2\leftrightarrow\bar{z}_jz_j$. In this example, the metaplectic correction contributed only in the case of the number operator, appearing as the extra $-\frac{1}{2}$ term in Eq. (\ref{bosonzz}) (it can already be seen in the light of the analysis in Section 2, that this resolves the issues concerning the harmonic oscillator path integral in the coherent states basis). In the opposite point of view, it is considered \cite{geom2} that the correct quantum physics for the observable $|z|^2$ are provided by the operator $z\frac{\partial}{\partial{z}}+\frac{1}{2}$ and not by $z\frac{\partial}{\partial{z}}$. This has been based on arguments related to the zero point energy of the harmonic oscillator and the commutator algebra appearing after quantization. Nevertheless, no direct or mathematically robust argument could be given until now, regarding why the correction should be included. In this example we proved that the de-quantization procedure gives such a reason, as this correction was vital for the computationally exact mapping between the canonical and path integral quantization of this system.
	
	Due to the linearity of the map we can also deduce the correct Hamiltonian symbol corresponding to any linear function of the previous three operators and the identity
	\begin{equation}
	\hat{F}(k,l,m,d)=k\hat{a}^\dagger\hat{a}+l\hat{a}+m\hat{a}^\dagger+d\mathbb{1},
	\end{equation}
	where $k,l,m,d\in\mathbb{C}$. This is found to be
	\begin{equation}
	\hat{Q}^{-1}\left(\hat{F}(k,l,m,d)\right)=k\left(|z|^2-\frac{1}{2}\right)+lz+m\bar{z}+d.
	\end{equation}
	As expected, if the metaplectic correction is not taken into account, these results reduce to the expectation values of the respective operators in the coherent states basis, which if used in time-continuous path integration give wrong results \cite{problems7_breakdown, kara1, vskohetov}, as explained in section 2.
	
	\subsection{Spin coherent states}
	
	For the spin coherent states (\ref{cospin}), the corresponding construction defines the kinetic term \cite{pathspinchoerent1, pathspincoherent2}
	\begin{equation}
	A_\mu^{(s)}\dot{x}^\mu=is\frac{(\bar{z}\dot{z}-z\dot{\bar{z}})}{1+|z|^2}, \label{goodcon}
	\end{equation}
	which yields the connection  ${A}^{(s)}=s\frac{i}{1+|z|^2}(\bar{z}dz-zd\bar{z})$. 
	The coordinate induced form of the $su(2)$ generator $\hat{S}_z$ in the highest weight s representation, can be found from its action on the coherent states as
	\begin{equation}
	\hat{S}_z\ket{z}_s^{(0)}{=}\left[-z\frac{\partial}{\partial{{z}}}+s\right]\ket{z}_s^{(0)}.
	\end{equation}
	For this operator, now, we can use Eq. (\ref{operator}) to compute the corresponding classical Hamiltonian.
	\\
	Plugging $\hat{S}_z$ in Eq. (\ref{operator}) one can find
	\begin{equation}
	\hat{Q}^{-1}\left(\hat{S}_z\right)=s\frac{1-|z|^2}{1+|z|^2}+\frac{1}{2}. \label{spinz}
	\end{equation}
	After this result, the partition function for the simple system $\hat{H}=\omega\hat{S}_z$
	assumes the
	following form
	\begin{align}
	Z=\text{Tr}\left[e^{-iT\omega\hat{S}_z}\right]=\mathcal{N}_S\underset{pbc}{\int}\left\{\prod_{t\in[0,T]}{2si}\frac{d^2z(t)}{(1+|z(t)|^2)^2}\right\} \label{examplespin}e^{i\int_{0}^{T}dt\left\{is\frac{\bar{z}(t)\dot{z}(t)-z(t)\dot{\bar{z}}(t)}{1+|z(t)|^2}-\omega\left(s\frac{1-|z(t)|^2}{1+|z(t)|^2}+\frac{1}{2}\right)\right\}}.
	\end{align}
	The measure is constructed such that
	\begin{align}
	\begin{split}
	\mathcal{N}_s\int\left\{\prod_{t\in[0,T]}{2si}\frac{d^2z(t)}{(1+|z(t)|^2)^2}\right\}
	=\underset{N\rightarrow\infty}{\text{lim}}\prod_{j=0}^{N}\frac{2s+1}{2\pi{i}}\int\frac{d^2z_j}{(1+|z_j|^2)^2} 
	\end{split}
	\end{align}
	and gives the canonical measure of integration coming from the coherent states.
	Note that, as in the bosonic case, the discrete form of the integral (\ref{examplespin}) is defined
	through the symmetric slicing $|z|^2\leftrightarrow\bar{z}_jz_j$. The integration in Eq. (\ref{examplespin}) is easily performed
	\cite{kara2} yielding the correct result $Z=\sum_{j=-s}^{s}e^{-i\omega{T}j}=\text{sin}[\omega{T}(s+1/2)]/\text{sin}[\omega{T}/2]$. The metaplectic correction, appearing as the $+\frac{1}{2}$ term in Eq. (\ref{spinz}), was once again exactly the term needed for the correct partition function to be recovered. 
	\\
	
	From Eqs. (\ref{bosonz}), (\ref{bosonzz}) and (\ref{spinz}) we deduce that if the classical observable of a quantum operator corresponds to a polarization preserving vector field, regardless if the metaplectic correction is taken into account, then the metaplectically corrected observable can be used as the Hamiltonian term in the corresponding time-continuous path integral. In the context of path integral quantization thus, the metaplectic correction appears to be a consistent and immediate way to arrive at correct results in the continuum, making its importance unambiguous. 
	
	At this point a comment is in order. It is interesting to see that the study of the resolution of the identity on the 2-Sphere, would give through the use of Eq. (\ref{identity}) 
	\begin{equation}
	\hat{I}_{S^2}=k'\int_{\mathbb{C}}\ket{z}\bra{z}e^{-Y_{S^2}(z,\bar{z})}\frac{dz\wedge{d\bar{z}}}{1+|z|^2}.
	\end{equation}
	If we demand this expression to coincide with the one appearing in Eq. (\ref{identspin}), the function in the exponent should be $Y_{S^2}=2\left(s+\frac{1}{2}\right)\text{ln}\left(1+|z|^2\right)$. If we proceed with geometric quantization at this level of structure, i.e. during the construction of the functional measure, the natural choice is to identify $Y_{S^2}$ as above. Taking into account that $Y_{S^2}$ is also defined through the solution of Eq. (\ref{polar}), we get, for the underlying symplectic structure, the 1-form ${A}'=\left(s+\frac{1}{2}\right)\frac{i}{1+|z|^2}(\bar{z}dz-zd\bar{z})$ which differs from the connection $A$ that appears in Eq. (\ref{goodcon}). As can be readily checked, the use of $A'$ instead of $A$ in Eq. (\ref{operator}) yields a Hamiltonian function which leads to a wrong result for the partition function (\ref{examplespin}). As a consequence, the symplectic structure indicated by the functional measure is different from the one indicated by the kinetic term of the action in this instance. This is not the case with the bosonic path integrals, where they coincide. Nevertheless, in the procedure proposed in this paper, the symplectic structure must be necessarily defined through the kinetic term, since it is immediately related to the classical mechanics defined by the action. \\
	
	Finally, note that spin systems with $s=1/2$ could also be addressed through fermionic path integrals, for which a method for the identification of the correct continuum limit has been proposed \cite{1st}.

\section{Higher orders and interactions}
\subsection{Higher orders}
The generalization of the previous results for some higher order operators can proceed with no need of additional mathematical structure, since the rigorous formalism of functional integrals allows for power reducing manipulations. Through these, path integration naturally defines a more general de-quantization map, which provides the proper Hamiltonian symbols for operators that are not necessarily $\hat{Q}^{-1}$-de-quantizable. By definition, this map is expected to share a lot of common traits with the inverse of a consistent - under Dirac's constraints - quantization map \cite{geom1,obstruction} (see Appendix A). Before we proceed, to avoid confusion and misconceptions, we stress that the map presented in the following is consistent in the path integral context, i.e. to acquire the classical Hamiltonian symbol pertaining to the action in a path integral. It is not guaranteed, though, to be an appropriate de-quantization scheme in general, and should not be used in contexts other than the path integral, unless, of course, it is properly justified.

So far, we have proposed a consistent de-quantization procedure which, through the action of $\hat{Q}^{-1}$, maps first order differential operators to their respective Hamiltonian symbols. Nevertheless, in order for path integration to be considered as a useful technique for actual systems, a generalization of this procedure to include higher order operators and interactions must be performed. We start by defining this procedure through the action of a more general de-quantization map $\mathcal{Q}^{-1}$, this time mapping arbitrary operators to their respective Hamiltonian symbols in the continuum 
\begin{equation}
\lim_{\epsilon\rightarrow0}\lim_{N\rightarrow\infty}H_{j}\equiv{}\mathcal{Q}^{-1}(\hat{H})(z(t),\bar{z}(t)). \label{truemap}
\end{equation} 
\\
The action of this map on $\hat{Q}^{-1}$-de-quantizable operators is defined as in the previous Section
\begin{equation}
\mathcal{Q}^{-1}|_{d.q.}=\hat{Q}^{-1},
\end{equation}
where $\hat{Q}^{-1}$ is the de-quantization map constructed via half-form quantization. In other words, the new map is identified with the map we introduced in Section 4, as far as $\hat{Q}^{-1}$-de-quantizable operators are concerned. The action of $\mathcal{Q}^{-1}$ on higher order operators cannot be understood this way though (at least not without introducing extra mathematical structure) and for this reason we will study it with the use of functional techniques. Note, that while in the definition of $\mathcal{Q}^{-1}$ we have used the exponent $-1$, in analogy to $\hat{Q}^{-1}$, it is not guaranteed that there exists a corresponding unique quantization map $\mathcal{Q}$ such that $(\mathcal{Q}^{-1})^{-1}=\mathcal{Q}$. Thus, we do not consider the map $\mathcal{Q}^{-1}$ as invertible; we merely use the $-1$ exponent to denote that it de-quantizes operators instead of quantizing functions.
In the same context we neither consider the map $\mathcal{Q}^{-1}$ to be necessarily linear on the space of operators, since during the usual slicing procedure the Hamiltonian symbol may acquire non-trivial contributions from commutator terms, in general. In the rest of this subsection we will study how $\mathcal{Q}^{-1}$ acts on polynomials of $\hat{Q}^{-1}$-de-quantizable operators. \\

In what follows, we consider time independent Hamiltonian operators and thus for simplicity we drop the time ordering operator $\hat{T}$ from the expressions. The functional identity we will use to identify the Hamiltonian symbol in higher orders is
\begin{align}
\begin{split}
\text{tr}\left[e^{\pm{}iT\hat{H}^2}\right]&\sim{}\text{tr}\left[\int\mathcal{D}\xi{e}^{\left\{\mp\frac{i}{4}\int_{0}^{T}dt\xi^2+i\int_{0}^{T}dt\xi{\hat{H}}\right\}}\right]=\int\mathcal{D}\xi{e}^{\mp\frac{i}{4}\int_{0}^{T}dt\xi^2}tr\left[{e}^{{i\int_{0}^{T}dt\xi\hat{H}}}\right]  \label{parathyraki2}
\end{split}
\end{align}
and its generalization for an arbitrary positive integer power $k$ 

	\begin{align}
	\begin{split}
	\text{tr}\left[e^{\pm{}iT\hat{H}^k}\right]&=\text{tr}\left[e^{\pm{}iT\left\{\frac{1}{2}\left(\hat{H}+\hat{H}^{k-1}\right)^2-\frac{1}{2}\hat{H}^2-\frac{1}{2}\hat{H}^{2k-2}\right\}}\right]\sim\\ &\sim\text{tr}\left[\int\mathcal{D}\xi_1\int\mathcal{D}\xi_2\int\mathcal{D}\xi_3e^{\mp\frac{i}{4}\int_{0}^{T}dt\left\{\xi_1^2-\xi_2^2-\xi_3^2\right\}}e^{\frac{i}{\sqrt{2}}\int_{0}^{T}dt\left(\xi_1+\xi_2\right)\hat{H}+\frac{i}{\sqrt{2}}\int_{0}^{T}dt\left(\xi_1+\xi_3\right)\hat{H}^{k-1}}\right],\label{hyperwindow}
	\end{split}
	\end{align}

both of which are valid for all operators $\hat{H}$ which have a complete set of eigenstates.
Through recursive use of Eq. (\ref{hyperwindow}), it is easy to prove the power mapping property
\begin{equation}
\mathcal{Q}^{-1}\left(\hat{H}^k\right)=\left(\hat{Q}^{-1}\left(\hat{H}\right)\right)^k, \text{ }k\in\mathbb{N}, \label{powermapping}
\end{equation}
for all time independent Hermitian $\hat{Q}^{-1}$-de-quantizable operators $\hat{H}$.
This provides the Hamiltonian symbol corresponding to the operator $\hat{H}^k$ in time-continuous coherent-state path integration. More explicitly, starting with $\text{tr}\left[e^{-iT\hat{H}^k}\right]$ and lowering the power of the operator to its first order, through the recursive use of Eq. (\ref{hyperwindow}), one can map the trace involving the first order operator to its path integral representation according to the results in Section 4.  Then, Eq. (\ref{powermapping}) is easily obtained, after integrating over the auxiliary fields $\xi_i$. This result can be generalized to include time dependent Hamiltonians, if the time dependence factors out as $\hat{H}(t)=f(t)\hat{H}$, where $\hat{H}$ is a constant Hermitian operator and $f(t)$ is a real smooth function.\\

For linear combinations of operators, even if all of them can be de-quantized, one cannot be sure that this can be done simultaneously for all of them, i.e. that $\mathcal{Q}^{-1}$ will act linearly on these. For a deeper understanding of such instances more sophisticated methods should be considered. Nevertheless, linearity is true when all operators appearing in the linear combination commute, since the previous construction can proceed independently for each operator. 
This can be confirmed by studying the action of $\mathcal{Q}^{-1}$ on the quantum operator
\begin{equation}
\hat{H}=\sum_{n=0}^{N}c_{n}(\hat{a}^{\dagger}\hat{a})^{n}, \quad{c_n}\in\mathbb{R}, \label{general}
\end{equation} 
the de-quantization of which, according to the previous arguments, gives
\begin{equation}
{H}_{cl}=\mathcal{Q}^{-1}\left(\sum_{n=0}^{N}c_{n}(\hat{a}^{\dagger}\hat{a})^{n}\right)=\sum_{n=0}^{N}c_{n}\left(|z|^2-\frac{1}{2}\right)^{n}.
\end{equation}
In cases like this, the factors $c_n$ can be considered time dependent, since the operators do not mix and continue to commute $\forall{}t\in[0,T]$. For the sake of simplicity though, we will continue by considering these as constants.
The calculation of the partition function
\begin{equation}
Z=\underset{Periodic}{\int}\mathcal{D}^2z(t) e^{i\int_{0}^{T}dt\left(\frac{i}{2}(\bar{z}\dot{z}-\dot{\bar{z}}z)-\sum_{n=0}^{N}c_{n}\left(|z|^2-\frac{1}{2}\right)^{n}\right)}, \label{integral}
\end{equation}
where\\ ${\int}\mathcal{D}^2z(t)=\mathcal{N}\int\left\{\prod_{t\in[0,T]}id^2z(t)\right\}=\underset{N\rightarrow\infty}{\text{lim}}\prod_{j=0}^{N}\frac{1}{2\pi{i}}\int{d^2z_j}$,
proceeds then by introducing the identity \cite{hubbardstratonovich1,hubbardstratonovich2,hubbardstratonovich3,hubbardstratonovich4} 
\begin{equation}
1=\int \mathcal{D}\zeta \delta[\zeta -|z|^{2}]=\int \mathcal{D}\zeta\int \mathcal{D}\sigma e^{-i\int_{0}^{T}dt\sigma(\zeta-|z|^2)}
\end{equation}
in Eq. (\ref{integral}).
As a result, the partition function (\ref{integral}) can be recasted into the form
\begin{align}
Z=\int\mathcal{D}\zeta\int \mathcal{D}\sigma{}e^{-i\int_{0}^{T} dt \left(\sigma \zeta+\sum_{n=0}^{N}c_{n}\left(\zeta -\frac{1}{2}\right)^{n}\right)}F(\zeta,\sigma),
\end{align}
where 
\begin{equation}
F(\zeta,\sigma)=\underset{\text{Periodic}}{\int}\mathcal{D}^2z {e}^{i\int_{0}^{T}dt\left(\frac{i}{2}\left(\bar{z}\dot{z}-\dot{\bar{z}}z\right)+\sigma|z|^2\right)}.
\end{equation}
The last integral can be calculated by standard means \cite{kleinert}, giving
$F(\zeta,\sigma)=\sum_{m=0}^{\infty}e^{i\int_{0}^{T}dt\sigma\left(m+\frac{1}{2}\right)}$ and consequently
\begin{equation}
Z=\sum_{m=0}^{\infty}\int \mathcal{D}\zeta \int\mathcal{D}\sigma e^{-i\int_{0}^{T} dt \left(\sigma \left(\zeta-m-\frac{1}{2}\right)+\sum_{n=0}^{N}c_{n}\left(\zeta -\frac{1}{2}\right)^{n}\right)}.
\end{equation} 
The integration over the field  $\sigma$ yields the functional delta $\delta[\zeta-m-\frac{1}{2}]$ and thus fixes the $\zeta$ variable to the values $m+\frac{1}{2}$, leading to
\begin{equation}
Z=\sum_{m=0}^{\infty}e^{-iT\sum_{n=0}^{N}c_{n}m^{n}},
\end{equation}
which is the correct result, as can be easily seen from the trace definition of $Z$. 

To study general Hamiltonian operators that are products of equal powers of $\hat{a}$ and $\hat{a}^\dagger$ or linear combinations of such terms, the correct mapping can be identified by exploiting the bosonic commutation relation to rewrite the Hamiltonian in the form of Eq. (\ref{general}), which in turn can be de-quantized as showcased above. This argument can be further generalized, to state that whenever a quantum Hamiltonian can be expressed as a polynomial of some Hermitian $\hat{Q}^{-1}$-de-quantizable operator, its mapping to the classical Hamiltonian pertaining to the path integral can be performed linearly using Eq. (\ref{powermapping}). In the case of bosonic systems, this argument can thus be used to study higher powers of the general Hermitian operator $\hat{H}=k\hat{a}^\dagger\hat{a}+c\hat{a}+\bar{c}\hat{a}^\dagger+d$, where $k,d\in\mathbb{R}$ and $c\in\mathbb{C}$, which is indeed a $\hat{Q}^{-1}$-de-quantizable operator.

Summarizing the results up to this point, for the bosonic case we have identified the action of the de-quantization map $\mathcal{Q}^{-1}$ on operators of the form
\begin{equation}
\hat{F}=\sum_{n=0}^{N}l_n\left(k\hat{a}^\dagger\hat{a}+c\hat{a}+\bar{c}\hat{a}^\dagger+d\right)^{n}, \label{bosquant}
\end{equation}
to be
\begin{equation}
\mathcal{Q}^{-1}\left(\hat{F}\right)=\sum_{n=0}^{N}l_n\left[k\left(|z|^2-\frac{1}{2}\right)+cz+\bar{c}\bar{z}+d\right]^{n}, \label{bosclas}
\end{equation}
providing the corresponding Hamiltonian symbols in the continuum. Here, $k,d\in\mathbb{R}$ and $c\in\mathbb{C}$ are constants, but the $l_n\in\mathbb{R}$  factors can in general be time dependent $\forall{n=1,\dots,N}$. More formally, we have shown that for constant $k,d\in\mathbb{R}$ and $c\in\mathbb{C}$, the map $\mathcal{Q}^{-1}$ takes elements linearly, from the set 
\begin{equation}
N_q^{(b)}=\text{span}_{\mathbb{R}}\left\{\left(k\hat{a}^\dagger\hat{a}+c\hat{a}+\bar{c}\hat{a}^\dagger+d\right)^n|n\geq0\right\} \label{bosequant1}
\end{equation}
to elements of the set
\begin{equation}
N_{cl}^{(b)}=\text{span}_{\mathbb{R}}\left\{\left[k\left(|z|^2-\frac{1}{2}\right)+cz+\bar{c}\bar{z}+d\right]^n\big|n\geq0\right\}.\label{bosecl1}
\end{equation}
When acting on the subset of operators (\ref{bosequant1}), $\mathcal{Q}^{-1}$ can be seen to be invertible. Furthermore, its inverse shares a lot of similarities with a quantization map subject to Dirac's constraints and as long as it only connects (\ref{bosequant1}) with (\ref{bosecl1}), it does not violate any no-go theorems presented in quantization theory \cite{geom1,obstruction}. At first glance, this might seem to contradict the results presented in Appendix A, but note that the set (\ref{bosecl1}) does not contain elements $z^k\bar{z}^l$ for any possible combination of integers $k$, $l$; it merely contains a specific subset of them, tailored to avoid obstructions. It must, also, be noted, that while operators belonging to (\ref{bosequant1}) span a relatively restricted set, they correspond to a large class of physical systems, many of which are very important. Indeed, the set (\ref{bosequant1}) encompasses, for example, any system whose Hamiltonian is a polynomial with respect to the number operator, making the path integral formalism applicable to such systems.\\

Considering spin systems, the classical Hamiltonian corresponding to the generator $\hat{S}_{z}$ in the highest weight $s$ representation of $su(2)$ appears in Eq. (\ref{spinz}). The de-quantization of its higher powers can be identified again, through the recursive use of Eq. (\ref{hyperwindow}).
In the functional integration formalism we can then identify the action of $\mathcal{Q}^{-1}$, this time taking elements linearly from the set
\begin{equation}
N_q^{(s)}=\text{span}_{\mathbb{R}}\left\{\hat{S}_z^n|n\geq0\right\} \label{spinquant}
\end{equation}
to elements of the set
\begin{equation}
N_{cl}^{(s)}=\text{span}_{\mathbb{R}}\left\{\left(s\frac{1-|z|^2}{1+|z|^2}+\frac{1}{2}\right)^n\big|n\geq0\right\},\label{spinclas}
\end{equation}
providing the correct Hamiltonian symbols. The validity of these symbols will be checked through a highly non-trivial example in the next subsection.
As expected, acting on the subset of operators (\ref{spinquant}), $\mathcal{Q}^{-1}$ is invertible and its inverse acting on the corresponding subset of functions (\ref{spinclas}) obeys Dirac's constraints \cite{geom1,obstruction}, since the algebra in both subsets is Abelian.\\

The aforementioned compatibility of the inverse of $\mathcal{Q}^{-1}$ with Dirac's constraints was, actually, anticipated due to the method used for its construction. To clarify this statement, we emphasize on some important steps in our procedure. At the level of algebra generators, the identification of the Hamiltonian symbol proceeded through the inversion of Eq. (\ref{operator}), where for the quantum operator $\hat{Q}$ the following property holds
\begin{equation}
[\hat{Q}(f),\hat{Q}(g)]=i\hat{Q}(\{f,g\}), \label{constraint}
\end{equation}
for all the consistently de-quantizable cases studied, i.e. for the operators $\hat{Q}=\hat{a}$, $\hat{a}^\dagger$, $\hat{a}^\dagger\hat{a}$ and $\hat{S}_z$.
This property is one of Dirac's constraints for the canonical quantization of classical observables. It is known that Eq. (\ref{constraint}) cannot hold for general sets of quantizable classical functions $\{f,g,\dots\}$,  since the inclusion of higher order classical observables in many occasions leads to inconsistencies \cite{geom1,obstruction,groenewolds2}. If the observables commute this property is trivially true, since both the Poisson brackets and the commutation relations of all elements in Eq. (\ref{constraint}) are zero.
In our case, the operator/function subsets (\ref{bosequant1})/(\ref{bosecl1}) and  (\ref{spinquant})/(\ref{spinclas}), connected through the general de-quantization map $\mathcal{Q}^{-1}$, arose by solving Eq. (\ref{operator}) for the classical observable $f$ -i.e. the correspondence $\hat{{Q}}(f)$/$f$ was obtained- and the resulting correspondence was generalized to higher powers through functional methods. This procedure produced Abelian subsets on which $\mathcal{Q}^{-1}$ is invertible and $\left(\mathcal{Q}^{-1}\right)^{-1}$ satisfies Eq. (\ref{constraint}), making it compatible with Dirac's constraints.
Nevertheless, even though geometric quantization provided an invertible map for the above examples, $\mathcal{Q}^{-1}$ may not be invertible in general and may bear no relation to an inverse theory of quantization.

\subsection{Interactions}

With the previously proposed methods, it is also possible to study some very general classes of interactions analytically. These are the cases involving linear combinations of tensor products of operators which can be simultaneously de-quantized and used as Hamiltonian symbols, weighing functional integrals. Such systems appear in the form
\begin{equation}
\hat{\mathcal{O}}_{int}=\sum_{l=1}^{N}\omega_{l}\hat{\mathcal{O}}^{l1}\otimes\hat{\mathcal{O}}^{l2}\otimes\cdots\otimes\hat{\mathcal{O}}^{lk}, \label{interaction}
\end{equation}
where $\hat{\mathcal{O}}^{lj}$ represents the operator of the j-th system participating in the $l$-th interaction term. 

To approach such systems, we first need to understand how the map $\mathcal{Q}^{-1}$ acts on tensor products of operators. Any consistent path integration map, providing the Hamiltonian symbols in the continuum limit, should map operators acting on different Hilbert spaces independently.  This property can be derived naturally from the discretization procedure. By definition then, $\mathcal{Q}^{-1}$ has this property which quantitatively is expressed as
\begin{equation}
\mathcal{Q}^{-1}\left({\hat{H}_1\otimes\hat{H}_2}\right)=\mathcal{Q}^{-1}\left({\hat{H}}_1\right)\mathcal{Q}^{-1}\left({\hat{H}}_2\right).
\end{equation}

Then, regarding operators of the form (\ref{interaction}), the map $\mathcal{Q}^{-1}$ acts on Eq. (\ref{interaction}) as
\begin{equation}
\mathcal{Q}^{-1}\left(\hat{\mathcal{O}}_{int}\right)=\sum_{l=1}^{N}\omega_{l}\mathcal{Q}^{-1}\left(\hat{\mathcal{O}}^{l1}\right)\mathcal{Q}^{-1}\left(\hat{\mathcal{O}}^{l2}\right)\cdots\mathcal{Q}^{-1}\left(\hat{\mathcal{O}}^{lk}\right), \label{intmap}
\end{equation}
as long as all $N$ operators corresponding to each subsystem can be simultaneously de-quantized.
Of course, this is trivially true when all $N$ operators commute.
In bosonic systems, the previous arguments allow for the consistent study of interaction terms between k subsystems such as:
\begin{enumerate}
	\item Interactions where all subsystems take part through a single first order operator and its powers:
	\begin{align}
	\begin{split}
	&\hat{\mathcal{O}}_{int{1}}^{(bos)}=\sum_{l=1}^{N}\omega_l\otimes_{j=1}^{k}\left(k_j\hat{a}^\dagger_j\hat{a}_j+c_j\hat{a}_j+\bar{c}_j\hat{a}^\dagger_j+d_j\right)^{n_{lj}},\phantom{000}\omega_l,k_j,d_j\in\mathbb{R},\quad{c}_j\in\mathbb{C}.\label{intbos}
	\end{split}
	\end{align}
	Here, the index $j$ refers to the subsystems, the index $l$ to the interaction term and the index $n_{lj}\in\mathbb{N}$ is the power of the $j$-th subsystem operator in the $l$-th interaction term.
	
	\item Interactions where at least one subsystem takes part through different, but simultaneously de-quantizable operators: 
	\begin{align}
	\begin{split}
	&\hat{\mathcal{O}}_{int{2}}^{(bos)}=F^{(1)}\otimes\hat{a}_j^\dagger\hat{a}_j+F^{(2)}\otimes\hat{a}_j+F^{(2)*}\otimes\hat{a}_j^\dagger+F^{(3)}\otimes\hat{I}_j.\label{intbos2}
	\end{split}
	\end{align}
	Here, the index $j$ refers to the subsystem with the aforementioned property, and $F^{(1)},F^{(3)}\in\mathbb{R}$, $F^{(2)}\in\mathbb{C}$ are functions of the operators of all the other subsystems, which also have to be simultaneously de-quantizable.
\end{enumerate}
In the same fashion, in spin systems we can study interactions of the form
\begin{equation}
\hat{\mathcal{O}}_{int}^{(spin)}=\sum_{l=1}^{N}\omega_l{\hat{S}_{z_1}}^{n_{l1}}\otimes{\hat{S}_{z_2}}^{n_{l2}}\otimes\cdots\otimes{\hat{S}_{z_k}}^{n_{lk}},\quad n_{lj}\in\mathbb{N}, \label{teleytaiostyp}
\end{equation}
where the indices have the same meaning as in the aforementioned bosonic case.
In this context, the operator ${\hat{S}_{z_j}}^{m}$ identifies the $m$-th power of the spin operator $\hat{S}_{z}$ of the $j$-th subsystem.\\ 

To showcase the validity of this mapping, we will study the most general $N$-system interaction Hamiltonian containing $\hat{S}_z$ operators up to a power $k$
\begin{equation}
\hat{H}_k=\sum_{a_1=1}^{k_1}\dots\sum_{a_N=1}^{k_N}c_{a_1\dots{a}_N}\hat{S}_{z_1}^{a_1}\otimes\cdots\otimes\hat{S}_{z_N}^{a_N},
\end{equation}
where $a_j\in\mathbb{N}$, $\sum_{j=1}^{N}k_j=k$ and $c_{a_1\dots{a}_N}\in\mathbb{R}$.
For this operator, our mapping gives 

	\begin{align}
	\begin{split}
	{H}^{(s)}_{cl}\left(|z_{1}|,\dots,|z_{N}|\right)&=\mathcal{Q}^{-1}\left(\sum_{a_1=1}^{k_1}\dots\sum_{a_N=1}^{k_N}c_{a_1\dots{a}_N}\hat{S}_{z_1}^{a_1}\otimes\cdots\otimes\hat{S}_{z_N}^{a_N}\right)=\\
	&=\sum_{a_1=1}^{k_1}\dots\sum_{a_N=1}^{k_N}c_{a_1\dots{a}_N}\left(s\frac{1-|{z}_{1}|^2}{1+|z_{1}|^2}+\frac{1}{2}\right)^{a_1}\cdots\left(s\frac{1-|{z}_{N}|^2}{1+|z_{N}|^2}+\frac{1}{2}\right)^{a_N}.
	\end{split}
	\end{align}

The calculation of the partition function
\begin{align}
Z&=\underset{Periodic}{\int}\left(\prod_{i=1}^{N}\mathcal{D}^2\mu(z_i)\right)e^{i\sum_{i=1}^{N}\int_{0}^{T}dt\left\{is\frac{\bar{z}_i\dot{z}_i-z_i\dot{\bar{z}}_i}{1+|z_i|^2}-{H}^{(s)}_{cl}\left(|z_{1}|,\dots,|z_{N}|\right)\right\}}, \label{partising}
\end{align}
where ${\int}\mathcal{D}^2\mu(z)=\mathcal{N}_S{\int}\left\{\prod_{t\in[0,T]}{2si}\frac{d^2z(t)}{(1+|z(t)|^2)^2}\right\}=\underset{N\rightarrow\infty}{\text{lim}}\prod_{j=0}^{N}\frac{2s+1}{2\pi{i}}\int{\frac{d^2z_j}{(1+|z_j|^2)^2}}$,
proceeds again by introducing the identity \cite{hubbardstratonovich1,hubbardstratonovich2,hubbardstratonovich3,hubbardstratonovich4} 
\begin{align}
\begin{split}
1&=\int\left(\prod_{i=1}^{N}\mathcal{D}\zeta_i\right)\delta\left[\zeta_i-\left(s\frac{1-|{z}_{i}|^2}{1+|z_{i}|^2}+\frac{1}{2}\right)\right]=\\
&=\int\left(\prod_{i=1}^{N}\mathcal{D}\zeta_i\right)\int\left(\prod_{i=1}^{N}\mathcal{D}\sigma_i\right) e^{-i\int_{0}^{T}dt\sigma_i\left(\zeta_i-\left(s\frac{1-|{z}_{i}|^2}{1+|z_{i}|^2}+\frac{1}{2}\right)\right)}
\end{split}
\end{align}
in Eq. (\ref{partising}).
As a result, the partition function (\ref{partising}) can be recasted into the form
\begin{align}
\begin{split}
Z&=\int\left(\prod_{i=1}^{N}\mathcal{D}\zeta_i\right)\int\left(\prod_{i=1}^{N}\mathcal{D}\sigma_i\right)e^{-i\sum_{i=1}^{N}\int_{0}^{T} dt \left(\sigma_i \zeta_i+{H}^{(s)}_{cl}\left(\zeta_1,\dots,\zeta_N\right)\right)}\prod_{i=1}^{N}F(\zeta_i,\sigma_i),
\end{split}
\end{align}
where 
\begin{equation}
F(\zeta_i,\sigma_i)=\underset{Periodic}{\int}\mathcal{D}^2\mu(z_i)e^{i\int_{0}^{T}dt\left\{is\frac{\bar{z}_i\dot{z}_i-z_i\dot{\bar{z}}_i}{1+|z_i|^2}+\sigma_i\left(s\frac{1-|{z}_{i}|^2}{1+|z_{i}|^2}+\frac{1}{2}\right)\right\}}
\end{equation}
and
\begin{equation}
{H}^{(s)}_{cl}\left(\zeta_1,\dots,\zeta_N\right)=\sum_{a_1=1}^{k_1}\dots\sum_{a_N=1}^{k_N}c_{a_1\dots{a}_N}\zeta_1^{a_1}\cdots\zeta_N^{a_N}.
\end{equation}
The last integral can be calculated by standard means \cite{kara2,coherentpath9+identitycoherentidentity3+spincalc+continuumsolari1,continuumsolari2+spincalc,coherentidentity4+spincalc+extraphase} and the result reads
$F(\zeta_i,\sigma_i)=\sum_{m_i=-s}^{s}e^{im_i\int_{0}^{T}dt\sigma_i}$,
leading to
\begin{align}
Z&=\sum_{m_1=-s}^{s}\cdots\sum_{m_N=-s}^{s}\int\left(\prod_{i=1}^{N}\mathcal{D}\zeta_i\right)\int\left(\prod_{i=1}^{N}\mathcal{D}\sigma_i\right)\times\\
&\times\text{exp}\left[-i\sum_{i=1}^{N}\int_{0}^{T} dt \left(\sigma_i\left(\zeta_i-m_i\right) +{H}^{(s)}_{cl}\left(\zeta_1,\dots,\zeta_N\right)\right)\right]\nonumber.
\end{align}
Finally, the integrations over the fields  $\sigma_i$ yield $N$ functional delta distributions $\delta\left[\zeta_i-m_i\right]$, fixing the $\zeta_i$ variables to the values $m_i$ and thus give
\begin{equation}
Z=\sum_{m_1=-s}^{s}\cdots\sum_{m_N=-s}^{s}e^{-iT\sum_{a_1=1}^{k_1}\dots\sum_{a_N=1}^{k_N}c_{a_1\dots{a}_N}m_1^{a_1}\cdots{m}_N^{a_N}},
\end{equation}
which is the correct result.\\

Note the physical significance of some of the mappings established in this Subsection; given a consistent and mathematically robust way to construct the continuum version of the path integral concerning Hamiltonians with interractions, a new tool becomes, now, available for the study of involved quantum systems, for which the conventional description of quantum mechanics in terms of operators is rather inconvenient. Indeed, our findings can be employed for the analytical study of interactions between systems (e.g. a qubit and its environment), both in real and imaginary time, only this time in terms of classical functions, and, thus, one can benefit from the convenience such a formalism provides.

\section{Conclusions}

Through the inversion of the half-form quantization procedure, we proposed a method for the consistent identification of Hamiltonian symbols in the continuum version of path integrals involving coherent states. This procedure was made formal through the construction of a de-quantization map taking operators to their corresponding Hamiltonian functions, weighing the aforementioned integrals, and was further generalized to higher orders through functional methods. The inclusion of the half-form structure was vital, since the metaplectic correction proved to be the missing ingredient in repairing the inconsistent results encountered in literature. 

Although the use of power reducing functional techniques made it possible to probe higher order operators, which can be seen not to preserve the polarization, the set of these operators may not be maximal and could  be extended in the future. This could be done either directly, by introducing extra mathematical structure in the construction of the de-quantization map, or indirectly, through the use of more sophisticated power reducing techniques. If the general form of the de-quantization map $\mathcal{Q}^{-1}$ is explicitly constructed, the most important mathematical direction would then be to understand how the de-quantization procedure, providing the correct Hamiltonian symbols in the continuum, differs from an inverse theory of quantization. An answer to this question may shed some light to the issues arising in the context of quantization theory. 

Even if such a task proves to be infeasible, though, with the aid of the de-quantization map proposed in the main text, physics have already acquired a new tool to tackle very important problems. Even the restricted set of operators we studied in the main text has a great physical significance, since such operators appear in a huge variety of important physical systems. Our findings, thus, can be used to study -both analytically and numerically- problems whose path integral could not even be defined consistently, until now, e.g. any polynomial of the number operator, system-environment interactions etc. Our findings are expected to be a helpful tool in fields where path integrals could have an important role, such as statistical physics and quantum information theory, among others. 

\section{Acknowledgements}
This research is co-financed by Greece and the European Union (European Social Fund- ESF) through the Operational Programme <<Human Resources Development, Education and Lifelong Learning>> in the context of the project ''Strengthening Human Resources Research Potential via Doctorate Research'' (MIS-5000432), implemented by the State Scholarships Foundation (IKY)

\appendix
\newpage

\section{Obstruction results in quantization theory}

Along the lines of quantization theory, a great number of no-go theorems is known, defining very strict bounds for the subsets of observables which can consistently be mapped to operators. In this context, consistency is defined as the compatibility of a mapping under the maximum possible number of Dirac's quantization constraints. Let $\hat{Q}:C^\infty(M)\rightarrow{End[H]}$ be the quantization map, taking functions lying on the phase space to operators acting on a Hilbert space. Then, Dirac's quantization constraints on $\hat{Q}$ have the form:
\begin{enumerate}
	\item $\mathbb{C}$-linearity $\hat{Q}(rf+g)=r\hat{Q}(f)+\hat{Q}(g)$,\\
	$r\in\mathbb{C}$, ${f,g}\in{C}^\infty(M)$.
	\item $\hat{Q}(1)=\hat{I}$.
	\item Hermitianicity with respect to the symplectic structure
	\begin{equation}
	\int{d^{2n}}q\sqrt{\omega}\Phi_1^*[\hat{Q}(f)\Phi_2]=\int{d^{2n}}q\sqrt{\omega}[\hat{Q}(f)\Phi_1]^*\Phi_2.
	\end{equation}
	\item $\hat{Q}$ defines a Lie-algebra homomorphism
	\begin{equation}
	[\hat{Q}(f),\hat{Q}(g)]=i\hat{Q}(\{f,g\}). \quad \label{homeo} 
	\end{equation}
	\item If $\{f_1,f_2,\dots,f_n\}$ is a complete set of observables then $\{\hat{Q}(f_1),\hat{Q}(f_2),\dots,\hat{Q}(f_n)\}$ is a complete set of operators.
\end{enumerate}
Unfortunately, it is impossible for a mapping to obey all of these constraints simultaneously, with the last two usually being weakened.  Groenewold-Van Hove theorem for the 2D Euclidean plane and its generalization on the 2-Sphere give the two simplest cases of this obstruction.
\begin{theorem}[Groenewold-Van Hove theorem]
	There does not exist a consistent quantization map $\hat{Q}$, which maps the position $(x)$ and momentum $(p)$ observables to their respective operators and at the same time holds for polynomials of degree equal or higher than three, with respect to $q$ and $p$ functions.
\end{theorem}
The proof for this theorem is very simple and can be summed up in only a few lines. Let $\hat{Q}$ be a homomorphism between the algebras of classical observables and operators - as indicated in (\ref{homeo}) - acting on the space of smooth functions in the usual way: $\hat{Q}(1)=\hat{I}$, $\hat{Q}(x)=\hat{x}$ and $\hat{Q}(p)=\hat{p}$. Exploiting (\ref{homeo}), one can identify the action of this map on second order operators
\begin{equation}
\hat{Q}(x^2)=\hat{x}^2,\quad \hat{Q}(p^2)=\hat{p}^2,\quad \hat{Q}(xp)=\frac{1}{2}(\hat{x}\hat{p}+\hat{p}\hat{x}),
\end{equation}
which is found to be unique.
Up to this point the subsets of observables and their operator images are closed and thus $\hat{Q}$ is indeed a homomorphism. On the contrary, the extension of this subset through the inclusion of higher order observables fails to remain consistent. To show this, firstly one has to extend the previous calculation and calculate
\begin{align}
\begin{split}
&\hat{Q}(x^3)=\hat{x}^3,\quad \hat{Q}(p^3)=\hat{p}^3,\\ &\hat{Q}(x^2p)=\frac{1}{2}(\hat{x}^2\hat{p}+\hat{p}\hat{x}^2), \quad \hat{Q}(xp^2)=\frac{1}{2}(\hat{x}\hat{p}^2+\hat{p}^2\hat{x}).
\end{split}
\end{align}
These results do not behave well under the closedness condition, since the calculation of $\{x^3,p^3\}$ and $\{x^2p,xp^2\}$ through (\ref{homeo}) gives rise to a false equality
\begin{equation}
\frac{1}{3}[\hat{x}^{3}, \hat{p}^{3}]=\frac{1}{4}[\hat{x}^{2}\hat{p}+\hat{p}\hat{x}^{2}, \hat{x}\hat{p}^{2}+\hat{p}^{2}\hat{x}].
\end{equation}
This result renders the higher order generalization of the map invalid.
Thus, the maximal quantizable subalgebra of the space of observables, containing the subset $\{1,p,x\}$, is $\{1,x,p,x^2,p^2,xp\}$ with $\hat{Q}$ defined from the previous mappings.

Through the canonical transformation
\begin{equation}
z=\frac{1}{\sqrt{2}}(x+ip), \quad \bar{z}=\frac{1}{\sqrt{2}}(x-ip),
\end{equation}
this can be rephrased, such that the maximal quantizable subalgebra of observables containing $\{1,z,\bar{z}\}$ is $\{1,z,\bar{z},z^2,\bar{z}^2,\bar{z}z\}$, mapping 1-1 to the subset of operators
\begin{equation}
\left\{1,\hat{a},\hat{a}^\dagger,\hat{a}^2,\hat{a}^{\dagger2},\hat{a}^\dagger\hat{a}+\frac{1}{2}\right\}. \label{wronge}
\end{equation}  
We also present the generalization of Groenewold's theorem for $S^2$ without the proof, since it is highly more involved\cite{groenewolds2}:
\begin{theorem}[Groenewold's theorem for $S^2$]
	Let $S_1,S_2,S_3$ be observable functions on the $2$-Sphere satisfying $\{S_i, S_j\}=\sum_{k=1}^{3}\epsilon_{ijk}{}S_k$. Then, the maximal Poisson subalgebra on the 2-Sphere, which contains  $\{1,S_1,S_2,S_3\}$, 
	and can be consistently quantized, is just that generated by $\{1,S_1,S_2,S_3\}$ itself.
\end{theorem}

We stress the fact, that what both of these Theorems state, is that an obstruction appears upon trying to quantize the set of all classical observables of a specific order -in Theorem 1 this happens during the quantization of third order functions, in Theorem 2 during the quantization of second order ones. This does not forbid, though, the existence of a consistent mapping between specific subsets of functions (possibly different than the ones presented above), tailored to avoid the obstructions that arise, and operators. Such subsets are (\ref{bosecl1})/(\ref{bosequant1}) and (\ref{spinclas})/(\ref{spinquant}) for which we were interested in due to their relevance to important physical problems. As explained in the main text, these subsets manage to bypass the obstruction and are compatible with Dirac's constraints, albeit they are not the ones encountered on the above Theorems. 

\section{Geometric Quantization and quantum operator equation}

In this Appendix we present the basics of geometric quantization theory, as it was used in this work and clarify our conventions. Again, we try to describe the significance of the objects that arise in our context.

We start by defining a symplectic manifold as an even dimensional differentiable manifold, on which a non-degenerate closed 2-form $\omega$ can be defined. Closed means $d\omega=0$, where $d$ is the exterior derivative -the extension of the usual differentiation relevant to differential forms; simply put, it raises the degree of a form by 1. Due to Poincare lemma it is known that such a closed quantity can always be locally expressed as 
\begin{equation}
\omega=-dA,
\end{equation}
where $A$ defines the symplectic potential. This has to do with the fact that the exterior derivative is nilpotent, i.e. $d^{2}f=0$. A vector field, which leaves $\omega$ invariant under its flow, is called a Hamiltonian vector field and can be related to a $C^\infty$-function $f$ through the equation
\begin{equation}
i_{\xi}\omega=-df,
\end{equation}
where $\xi$ is the Hamiltonian vector field and $i_{\xi}$ is the interior derivative -it lowers the degree of a differential form by 1. Specifically for 1-forms, such as the symplectic potential, we have the formula $i_{\xi}A=\xi^{\mu}A_{\mu}$, where the right hand side of this equation is the dual pairing between the vector field and the 1-form. 
Canonical transformations are those transformations that leave the symplectic 2-form invariant. 

Then, geometric quantization proceeds by defining for each smooth function $f$ the prequantum operator, acting on wave functions, as
\begin{equation}
\hat{P}(f)=-i(\xi^\mu{D}_\mu+if),
\end{equation}
where $D_\mu=\partial_\mu+iA_\mu$ is the covariant derivative.  

It is trivial to confirm that this operator satisfies the first two of Dirac's constraints. It, also, constitutes a Lie algebra homomorphism between the space of smooth functions and the space of operators acting on a Hilbert space, since 
\begin{equation}
[\hat{P}(f),\hat{P}(g)]=i\hat{P}(\{f,g\})
\end{equation}
and it is Hermitian under the canonical inner product of the symplectic manifold
\begin{equation}
(\Phi_1,\Phi_2)=\int_Md^{2n}x\sqrt{\text{det}\omega(x^\mu)}\Phi_1^*(x^\mu)\Phi_2(x^\mu),
\end{equation}
which confirms that it obeys the third and the fourth constraint, as well. This operator is called prequantum, though, because it defines reducible representations of quantum mechanics. This means that wave functions depend on all the coordinates of the phase space, which is not the case in conventional quantum mechanics. This needs to be fixed. To recover an irreducible representation one restricts the operator to a set of polarized wave-functions, where through a polarization, the phase space dimension is reduced by half.

The new measure of integration $J$, on the reduced phase space, though, is not guaranteed to share the same properties as the one induced from the symplectic structure and in the general case the Hermitianicity condition breaks down 
\begin{align}
\begin{split}
&\int{J}d^nx\Phi^*_1(\hat{P}(f)\Phi_2)-\int{J}d^nx(\hat{P}(f)\Phi_1)^*\Phi_2=i\int{J}d^nx[\partial_\mu\xi^\mu+\xi^\mu\partial_\mu\text{ln}J]\Phi_1^*\Phi_2.
\end{split}
\end{align}
This issue is bypassed by defining the wave functions as half-form elements
\begin{equation}
\Phi_j(x)\sqrt{J(x)}.
\end{equation}
We can then define a quantum operator $\hat{Q}$, which acts on these as 
\begin{align}
\begin{split}
&\hat{{Q}}(f)[\Phi(x)\sqrt{J(x)}]=[-i(\xi^\mu{D}_\mu+if)\Phi(x)]\sqrt{J(x)}-i\Phi(x)\mathcal{L}_{\xi}\sqrt{J(x)},
\end{split}
\end{align}
where
\begin{equation}
\mathcal{L}_{\xi}\sqrt{J(x)}=\left(\frac{1}{2}\partial_\mu\xi^\mu+\xi^\mu\partial_\mu\right)\sqrt{J(x)}
\end{equation}
is the Lie derivative acting on the half-form $\sqrt{J(x)}$ and the index $\mu$ labels now the coordinates of the reduced phase space.

On a 2D manifold, if holomorphic polarization is chosen, the half-form part of a holomorphically polarized wave-function (\ref{halfform}) can be chosen to be $\sqrt{dz}$, since any extra coefficient can be absorbed in the holomorphic function $\phi(z)$. This way we get
\begin{equation}
\mathcal{L}_{\xi}\sqrt{{dz}}=\frac{1}{2}\partial_z\xi^z\sqrt{dz}.
\end{equation}
The action of the operator on the holomorphically polarized wave-functions can then be expressed  as
\begin{equation}
\hat{{Q}}(f)[\Phi(z,\bar{z})]=-i\left(\xi^z{D}_z+if+\frac{1}{2}\partial_z\xi^z\right)\Phi(z,\bar{z}).
\end{equation}
Finally, by writing explicitly the form of the holomorphic wave-functions as
\begin{equation}
\Phi(z,\bar{z})\sqrt{dz}=\phi(z)e^{-Y(z,\bar{z})/2}\sqrt{dz}
\end{equation}
we can find the action of the quantum operator on the function $\phi(z)$ to be
\begin{equation}
\hat{Q}(f)\phi(z)=\left(-i\xi^{z}\partial_{z}-\frac{i}{2}\left(\partial_{z}+4iA_{z}\right)\xi^{z}+f\right)\phi(z). 
\end{equation}

\end{document}